# Data structure for representing a graph: combination of linked list and hash table

*Maxim A. Kolosovskiy*
*Altai State Technical University, Russia*
*maxim.astu@gmail.com*

***Abstract***

*In this article we discuss a data structure, which combines advantages of two different ways for representing graphs: adjacency matrix and collection of adjacency lists. This data structure can fast add and search edges (advantages of adjacency matrix), use linear amount of memory, let to obtain adjacency list for certain vertex (advantages of collection of adjacency lists). Basic knowledge of linked lists and hash tables is required to understand this article. The article contains examples of implementation on Java.*

# 1. Introduction

There are two most common ways for representing a graph:

- Adjacency matrix
- Collection of adjacency lists

Let's look at comparison table of these ways:

|  | Adjacency matrix | Collection of adjacency lists |
|---|---|---|
| Memory complexity *(optimal – O(|E|))* | $O(|V|^2)$ | **O(|E|)** |
| Add new edge *(optimal – O(1))* | **O(1)** | **O(1)** |
| Remove edge *(optimal – O(1))* | **O(1)** | $O(|K|)$ |
| Search edge *(optimal – O(1))* | **O(1)** | $O(|K|)$ |
| Enumeration of the vertices adjacent to u *(optimal – O(|K|))* | $O(|V|)$ | **O(|K|)** |

*Table 1. Comparison table of two ways for representing graphs. V – set of vertices, E – set of edges, K – set of vertices adjacent to vertex u. We assume that |V|<|E|. Bold font – optimal complexity.*

Operations with a graph represented by an adjacency matrix are faster. But if a graph is large we can't use such big matrix to represent a graph, so we should use collection of adjacency lists, which is more compact. Using adjacency lists is preferable, when a graph is sparse, i.e. |E| is much less than $|V|^2$, but if |E| is close to $|V|^2$, choose adjacency matrix, because in any case we should use $O(|V|^2)$ memory.

Adjacency matrix and adjacency lists can be used for both directed and undirected graphs. In this article we consider directed graphs.

In this article we discuss a data structure, which combines these two ways representing graphs. This structure is a union of two data structures: linked list and hash table. Let name of this structure be *HashList*. *HashList* has following memory and time complexities characteristics:

|  | HashList |
|---|---|
| Memory complexity *(optimal – O(|E|))* | **O(|E|)** |
| Add new edge *(optimal – O(1))* | **O(1)** |
| Remove edge *(optimal – O(1))* | – |
| Search edge *(optimal – O(1))* | **O(1)** |
| Enumeration of the vertices adjacent to u *(optimal – O(|K|))* | **O(|K|)** |

*Table 2. Memory and time complexities for HashList.*

Before *HashList*'s consideration we recall how we can use linked list and hash table for storing a graph.

## 2. Linked list for representing a graph

In this paragraph we consider the simplest implementation to represent a graph by a collection of adjacency lists. The main idea of this way is storing a linked list of adjacent vertices for each vertex.

Talk about implementation of linked list in common. The clearest way is to use object-oriented approach: for each element in list we create a new instance of specific class (let it be called *Item*). This class should contain follow fields:

- data (it is integer value for our task, this value keeps number of vertex)
- pointer to previous element in the list (this field is not necessary for our task)
- pointer to next element in the list

In main program pointer to head of list must be defined. If this pointer is null, list is empty.

This implementation can be improved by replacing classes by arrays with integer values. So we should define following arrays:

- array for storing data (call it *data*)
- array for storing pointers to next elements (call it *next*)

In main program we will define integer value *head*, which contains index of head of list in these arrays. Sizes of these arrays equal (number of edges in graph + 1), because value 0 in *next* means that there is no pointer to the next element (value 0 in variable *head* means that the list is empty), so we shouldn't use cell with index 0.

Come back to representation of graphs. We should make a change to use the previous implementation of linked list – we replace single integer value *head* by once more array (call it *heads*), which contains head of list of adjacent vertices for each vertex. Other arrays (*data* and *next*) remain common for all vertices. And one more note about using arrays instead class: we should know what cells in arrays are free and what are used, in our implementation we define variable *used*, which contains number of used cells and when we need to get a free cell we will take (*used* + 1).

Thus, we obtain implementation of data structure for representing graphs (call it *MultiList* – list with multiple heads). Java code for described implementation:

```java
class MultiList {
  int used;
  int[] heads, next, data;
  // n - number of vertices, m - number of edges
  MultiList(int n, int m) {
      heads = new int[n];
      next = new int[m + 1];
      data = new int[m + 1];
```

```
        }
        // appends new edge (x, y)
        void add(int x, int y) {
            used++;
            data[used] = y;
            next[used] = heads[x];
            heads[x] = used;
        }
        // returns true if edge (x, y) is contained in the graph
        boolean contains(int x, int y) {
            for (int i = heads[x]; i != 0; i = next[i])
                if (data[i] == y)
                    return true;
            return false;
        }
    }
```

## 3. Hash table for representing a graph

The reason of using hash table is quick search of edge in graphs: we can search edge with O(1) as well as in adjacency matrix.

There are some different kinds of implementation of hash table:

- direct-address
- chained
- open-addressing

We consider open-addressing hash table with linear probing for resolving collisions (method for resolving collision doesn't matter, we can use double hashing, quadratic probing). We can use chained hash table, but this implementation is more complicated.

We will use a hash table which can add and search elements, but can't remove elements, because it will cause some troubles with open-addressing hash table. So *HashList* will not remove edges too. But if we use table supporting removing, *HashList* can remove edges too, it is not a fatal restriction.

We can use hash table for representing a graph in the following way: the table contains all edges, so we can add new edges and search edge.

Java code of such implementation:

```
class HashTable {
    final int SIZE = 1000000; // table size
    long[] data = new long[SIZE];
    boolean[] used = new boolean[SIZE];
    // adds new edge (x, y)
    boolean add(int x, int y) {
```

```java
            long code = code(x, y);
            int hash = hash(x, y);
            while (used[hash])
                if (data[hash] == code)
                    return false;
                else
                    hash = (hash + 1) % SIZE;
            used[hash] = true;
            data[hash] = code;
            return true;
    }
    // returns true if edge (x, y) is contained in hash table
    boolean contains(int x, int y) {
            long code = code(x, y);
            int hash = hash(x, y);
            while (used[hash])
                if (data[hash] == code)
                    return true;
                else
                    hash = (hash + 1) % SIZE;
            return false;
    }
    // converts pair (x, y) to a single value
    long code(int x, int y) {
        return ((1L * x) << 32) | y;
    }
    // returns hash code of edge (x, y)
    int hash(int x, int y) {
        return Math.abs((x + 111111) * (y - 333333) % SIZE);
    }
}
```

Sizes of arrays *data* and *used* are determined by the number of edges in the graph and the desirable load-factor of hash table. Too small load-factor reduces the execution time of adding and searching elements in table and increases the amount of memory, and vice versa for too large load-factor.

## 4. Merging linked list and hash table

The goal of merging linked list with hash table is merging advantages of linked list (optimal enumeration of adjacency list of vertex) and hash table (fast searching and adding edges). To reach this goal we enlarge amount of memory, but save linear memory complexity $O(|E|)$.

To explain how *HashList* works we return to working of linked lists. If we insert new edge we get index of free cells in arrays *next* and *data*. For this purpose we use single integer variable *used*. In *HashList* we use other routine. We determine position for new edge like as we do it in hash table: we calculate hash code and in case of collision we resolve it (for example, by linear probing). So if we need to find edge we search it in certain place in arrays *next* and *data*.

Look at Java code implementing this idea:

```java
class HashList {
    final int SIZE = 1000000;
    int[] heads;
    long[] data = new long[SIZE];
    boolean[] used = new boolean[SIZE];
    int[] next = new int[SIZE];
    // n - number of vertices
    HashList(int n) {
         heads = new int[n];
         Arrays.fill(heads, -1);
    }
    // adds new edge (x, y)
    boolean add(int x, int y) {
         long code = code(x, y);
         int hash = hash(x, y);
         while (used[hash])
              if (data[hash] == code)
                   return false;
              else
                   hash = (hash + 1) % SIZE;
         data[hash] = code;
         used[hash] = true;
         next[hash] = heads[x];
         heads[x] = hash;
         return true;
    }
    // returns true if edge (x, y) is contained in the graph
    boolean contains(int x, int y) {
         long code = code(x, y);
         int hash = hash(x, y);
         while (used[hash])
              if (data[hash] == code)
                   return true;
              else
                   hash = (hash + 1) % SIZE;
         return false;
    }
    // enumerates the vertices adjacent to x
    void enumerate(int x) {
         for (int i = heads[x]; i != -1; i = next[i]) {
              int y = (int) data[i];
              // do something with y
         }
    }
    // returns hash code for edge (x, y)
    int hash(int x, int y) {
         return Math.abs((x + 111111) * (y - 333333) % SIZE);
    }
    // converts pair (x, y) to single integer value
    long code(int x, int y) {
         return ((1L * x) << 32) | y;
    }
}
```

Let's make analysis of memory and time complexities for linked list and hash table when we use these structures for representing a graph:

|  | Linked list | Hash table | HashList |
|---|---|---|---|
| Memory complexity *(optimal – O(|E|))* | O(|V|+|E|) | **O(|E|)** | **O(|E|)** |
| Add new edge *(optimal – O(1))* | **O(1)** | **O(1)** | **O(1)** |
| Remove edge *(optimal – O(1))* | O(|K|) | – | – |
| Search edge *(optimal – O(1))* | O(|K|) | **O(1)** | **O(1)** |
| Enumeration of the vertices adjacent to u *(optimal – O(|K|))* | **O(|K|)** | O(|V|) | **O(|K|)** |

*Table 3. Comparison table of memory and time complexities of HashList and its "components"*

This table shows that all characteristics of *HashList* are optimal.

## 5. Conclusion

In this article we obtained a data structure, which quickly performs basically operations with graphs and use optimal amount of memory. We made it by merging two separate data structures. We showed that it is possible to combine their functionality while saving optimal time and memory complexities.

This data structure can be used for storing different kind of graphs: trees, tries, weighted graphs, flow networks, multigraphs. For example, to store weighted graphs we should only add one array, which contains weight for each edge.